\newcommand\beq{\begin{eqnarray}}
\newcommand\eeq{\end{eqnarray}}

\newcommand\la{\langle}
\newcommand\ra{\rangle}
\newcommand{\cl}{\centerline}

\newcommand{\op}{{\cal O}}

\documentstyle[12pt]{article}
\begin{document}
\setlength{\baselineskip}{0.3in}
\setcounter{section}{0}
\setcounter{equation}{0}
\begin{titlepage}
\cl{\Large{\bf Twist-4 Matrix Elements of the Nucleon}}
\cl{\Large{\bf from Recent DIS Data at CERN and SLAC}}
\vspace{1in}
\cl{S. Choi$^1$, T. Hatsuda$^2$,
 Y. Koike$^3$  and Su H. Lee$^{1,2}$ }

\vspace{0.8cm}

$^1${\em Physics Department, Yonsei University, Seoul 120-749, Korea}

\vspace{0.3cm}

$^2${\em Physics Department, FM-15, University of Washington,
 Seattle, WA 98195, USA}

\vspace{0.3cm}

$^3${\em National Superconducting Cyclotron Laboratory,
Michigan State University,}

{\em  \  East Lansing, MI 48824-1321, USA}

\vspace{1in}
\cl{\bf Abstract}

We analyse the recent precision measurements of the lepton-hadron deep
inelastic scattering at CERN and SLAC to extract model independent
constraints among  the nucleon matrix elements of the twist-4
 operators.
 We also study a parameterization of these
 matrix elements and point out the possibility
 that
 the matrix elements of the quark-gluon mixed operator
has a negative value
 of the order of $- (400 \pm 100\ {\rm MeV})^2$ at 5 GeV$^2$
 renormalization scale.
\end{titlepage}

\section{Introduction}
\setcounter{equation}{0}

 Measurements of the lepton-hadron
 deep inelastic scattering(DIS)
 remain to be the cornerstone of various
QCD tests,
ranging from precise determination of $\Lambda_{QCD}$ to the
knowledge of the
structure functions necessary
to calculate cross sections for hard scattering
processes. (For the recent review, see \cite{NMC92}.)

 Recent precision DIS data  at CERN
 \cite{NMC1,NMC2,BCDMS} and at SLAC \cite{SLAC}
also provide us
 with a fruitful byproduct, i.e. the estimate of
 higher twist effects in the  spin averaged structure
 functions ($F_2$ and $F_L$).
 The twist-4 part of these structure functions
 is defined through
\beq
F_{2,L}(x,Q^2) = F_{2,L}^{\tau=2} (x,Q^2) + {1 \over Q^2} F_{2,L}^{\tau =4}
(x,Q^2) ,
\label{eq1}
\eeq
where the target mass corrections \cite{NGP} are taken into account
 in the twist-2 part $F_{2,L}^{\tau =2}$.

In terms of  the operator product expansion (OPE), the four-quark
 operators  ($\bar{\Psi} \Gamma_{\mu_1} \Psi
\bar{\Psi} \Gamma_{\mu_2} \Psi$) and the quark-gluon mixed
operator ($\bar{\Psi} \{D_{\mu_1},{}^*F_{\mu_2 \alpha}\} \Gamma^\alpha \Psi$)
 contribute to $F_{2,L}^{\tau =4}$ \cite{P80,SV82,JS,EFP82}.
 The twist-2 part $F_{2,L}^{\tau =2}$  is known to give a parton
distribution
 (i.e. the single particle property of quarks
and gluons in the nucleon),
 while the matrix elements of the twist-4 operators
  are the measure of the
 correlation of quarks and gluons in the nucleon.

Such new  information has wide applications in QCD:
  first of all it gives a detailed knowledge of
 the nucleon structure and gives a stringent test of the various
  models of the nucleon.
 Secondly, these twist-4 matrix elements are useful
 to analyse the higher twist
effects in other high energy processes such
 as the neutrino induced reaction and
the Drell-Yan processes \cite{QIU}.
  Thirdly,  the twist-4 matrix elements are
 essential to study the propagation of hadrons
  in nuclear medium as is shown in the framework of
  the QCD sum rules \cite{HKL}.

At present, an unambiguous determination of the magnitude of the
 twist-4 matrix elements is not available. However, the recent
  NMC data \cite{NMC1,NMC2} together with the
SLAC \cite{SLAC} and BCDMS \cite{BCDMS} data give us a useful
 constraint among the twist-4 matrix elements.  In this paper,
 we will first examine such constraints in a  model
independent way.
 Then, we will introduce a parameterization to satisfy the
 constraints and  point out that
the quark-gluon mixed operator at
 5 GeV$^2$ scale has a sizable nucleon
  matrix element.

\section{Operator Product Expansion}

The spin-2, twist-4 contribution to the spin-averaged
forward amplitude of the
electromagnetic current $j_{\mu}^{em}$
 can be written as \cite{SV82,JS}
\beq
T_{\mu \nu} & = & i \int d^4\xi\, e^{iq\xi} \la T j_\mu^{em}(\xi)
j_\nu^{em}(0) \ra_N
\nonumber \\
 & \rightarrow & d_{\mu \nu} \frac{1}{x^2 Q^2}
( A^1+\frac{5}{8} A^2+\frac{1}{16}
  A^g )  + e_{\mu \nu} \frac{1}{x^2 Q^2} ( \frac{1}{4} A^2 -
\frac{3}{8} A^g ) ,
\label{eq2}
\eeq
where the polarization tensors are defined as
 $e_{\mu \nu}   =  g_{\mu \nu} - q_\mu q_\nu /q^2 $ and
$d_{\mu \nu} =  -p_\mu p_\nu q^2/(p \cdot q)^2
+ (p_\mu q_\nu +p_\nu q_\mu)/p \cdot q -g_{\mu \nu}$ with
 $Q^2$=$-q^2$.  ($p^{\mu}$ is a 4-momentum of the nucleon
 with $p^2=M_N^2$.)
 $A^{1,2,g}$ are the spin-averaged nucleon matrix
elements of the spin-2,
 twist-4 operators:
\beq
 \la \op^k_{\alpha \beta} \ra= (p_\alpha p_\beta -
\frac{1}{4}M_N^2 g_{\alpha \beta})A^k,
\label{eq3}
\eeq
with
\beq
\op^1_{\alpha \beta} & = & g^2
  (\bar{q} \gamma_\alpha \gamma_5 {\cal Q} t^a q)(
\bar{q} \gamma_\beta \gamma_5 {\cal Q} t^a q ) ,\nonumber \\[12pt]
\op^2_{\alpha \beta} & = & g^2 ( \bar{q} \gamma_\alpha
 {\cal Q}^2 t^a q)(
\bar{q} \gamma_\beta   t^a q ), \nonumber \\[12pt]
\op^g_{\alpha \beta} & = & ig
  ( \bar{q} \{ D_\alpha , { }^*F_{\beta \mu} \}
 \gamma^\mu \gamma_5 {\cal Q}^2 q ) .
\label{eq4}
\eeq
Here,  the operators are assumed to be symmetric
and traceless with respect to the Lorentz indices:
 $\op_{\alpha \beta}
 \rightarrow {1 \over 2} (\op_{\alpha \beta}+\op_{\beta \alpha})
- {1 \over 4} g_{\alpha \beta} \op_{\gamma \gamma}$.
 ${\cal Q}$
is the  flavor $SU(2)$ charge matrix and
 $t^a$ are the generators of the
color SU(3) normalized to tr$(t^a)^2$=1/2.
$F_{\alpha \beta}=F_{\alpha \beta}^a t^a$, and
 the dual field strength is defined as $ ^*F_{\alpha \beta} =
\epsilon_{\alpha \beta \gamma \delta} F^{\gamma \delta}$ with
 $\epsilon_{0123}=1$.
 Here we have neglected the twist-4 operators
proportional to the current quark masses.
 A typical diagram which generates  $\op^1$ is given in Fig.\,1(a), and
 that for $\op^g,\op^2$
   is given in  Fig.\,1(b).   If one
 writes eq.\,(\ref{eq2}) as $T= 2 M /x^2 Q^2$, twist-4
matrix elements and  the twist-4
 structure functions are related as
\beq
M_{2,L} (Q^2) = \int_0^1 dx F_{2,L}^{\tau=4}(x,Q^2).
\label{eq5}
\eeq

\vspace{0.5cm}

\section{ Experimental data }

{\bf  Structure Function $F_2(x)$}

\vspace{0.2cm}

The experimental data of $F_2^{\tau =4}(x)$
 have been analyzed by introducing the following unknown function $C(x)$
\beq
F_2^{\tau=4}= C(x) F_2^{LT}(x,Q^2) ,
\label{eq6}
\eeq
where $F_2^{LT}(x,Q^2)$ denotes the  leading-twist
structure function with the target mass correction\,\cite{NGP}.

 $C(x)$ has been extracted for the hydrogen and
deuterium target
in ref. \cite{VM} by using the  BCDMS data and the SLAC data
 taken in the kinematic
region $0.07 < x < 0.75$ and $0.5 < Q^2 < 260 $ GeV$^2$.
 We have carried out
 $\chi^2$ fitting of the proton data $C_p(x)$ (given in Table 2
 of \cite{VM}) by
\beq
C(x) = a_0+a_1x + a_2x^2+ a_3 x^3 +a_4 x^4 \ \ \ ,
\label{eq7}
\eeq
 and we get  $a_0=  -0.28$,
$a_1=  3.45$,
$a_2= -17.13$,
$a_3=  31.64$,
and $a_4=  -14.95$.

 One can also  extract
$C_n(x)$ by combining hydrogen and
 deuterium data in \cite{VM}.
 The result,
however, has large error bars.
  On the other hand, the NMC group
recently published  better statistics
 data for $C_p(x)-C_n(x)$ (but not for
$C_p(x)$ and $C_n(x)$ separately)
  which is a combination of NMC, SLAC and BCDMS data \cite{NMC2}.
 The NMC group analyzed the
ratio $F^n_2/F^p_n$  in
the kinematic range $0.07 < x < 0.75$ and $0.8 < Q^2 < 75\ {\rm GeV}^2$.
   This ratio is independent
of the spectrometer acceptance and normalization and  gives a reliable
estimate of $C_p(x)-C_n(x)$ from
the following relation,
\beq
\frac{F^n_2}{F^p_2}= ( \frac{F^n_2}{F^p_2} )^{LT} ( 1-
\frac{C_p(x)-C_n(x)}{Q^2} ).
\label{eq8}
\eeq
By combining this data with that of the proton in ref. \cite{VM} and
fitting the resulting values for $C_n(x)$ with the same polynomial
in eq.\,(\ref{eq7}), we obtain the following values for the coefficients;
$a_0= -0.28$,
$a_1=   3.12$,
$a_2=  -11.01$,
$a_3=  16.51$,
 and $a_4=  -2.40$.
We  checked that different set of fittings fall well within the
estimated errors of the following results.

  From our fit of $C_p(x)$ and $C_n(x)$,
the integrated structure function
 at $Q^2=5\ {\rm GeV}^2$
 (which is a typical scale where the twist-4 effect
 is extracted) reads
\beq
\int_0^1 F_2^{\tau =4}dx & = & \frac{1}{2} (A^1 +\frac{5}{8}
A^2+\frac{1}{16}A^g  )
  \nonumber \\
& = &  \int_0^{1} C(x) F_2^{LT}(x)  dx  =  \left\{ \begin{array}{ll}
             0.005 \pm  0.004   &  {\rm GeV}^2 \ \ \mbox{(proton}) \\
             0.011 \pm  0.004 &  {\rm GeV}^2 \ \ \mbox{(neutron)}
                                \end{array} \right. \ .
\label{eq9}
\eeq
 The errors come from unavailability of  $C(x)$
for  $x>0.75$ and
 $x< 0.07$.   Here we have used the leading order (LO)
structure function
 of Gl\" {u}ck-Reya-Vogt \cite{GRV} for $F_2^{LT}$.
At $Q^2 \sim 5$\ GeV$^2$, the
 difference between the LO and the higher order (HO)
distribution functions are not
 significant after the $x$-integration.
\footnote{
Although this phenomenological parton distribution function might
contain the effect of the power corrections, this portion will be
$O(1/Q^4)$ and thus irrelevant in the twist-4 part of
$F_2$ in eq.\,(\ref{eq1}).}

\vspace{1cm}

\noindent{\bf Longitudinal Structure Function $F_L(x)$}

\vspace{0.2cm}

The higher twist effect in the longitudinal structure
function is obtained by the ratio between the longitudinal and
transverse cross sections $ R=\sigma_{_L}/\sigma_{_T}$ .
This ratio is especially sensitive
 to the higher twist contribution because the lowest
twist effect to $F_L$ is of order $\alpha_s$.
 Note that only diagrams such as
 given in Fig.\,1(b) contribute to $F_L$.  In
this case,
the twist-4 analysis using the transverse basis
provides us with an intuitive picture
\cite{EFP82},
in which the higher twist
effects can be interpreted in terms of the
 intrinsic transverse momentum
  of partons:
 $F_L^{\tau=4}(x) = 4 \int d^2 k_T k_T^2 f(x,k_T^2) $,
where $f(x,k_T^2)$ denotes a structure function for quarks with
the momentum fraction $x$ and the transverse momentum $k_T$.

Motivated by this, the SLAC data \cite{SLAC} was analysed
 in ref. \cite{MMS89} by introducing a typical scale for the transverse
momentum of the parton $\kappa$,\footnote{This  will be an
important guide  for our parameterization in section 3.}
\beq
F_L^{\tau=4}(x,Q^2)={8 \kappa^2} F_2^{LT}(x,Q^2).
\label{eq10}
\eeq
By using the leading and next-to-leading order
  structure function for $F_2^{LT}$,
 the SLAC data can be fitted by
\beq
\kappa^2=0.03 \pm 0.01 \ {\rm GeV}^2,
\label{eq11}
\eeq
in the range $0.2< x < 0.6$ \cite{MMS89,GMMPS91}.
An indirect experimental justification of eq.\,(\ref{eq10}) is that $R$
is independent of targets \cite{NMC1,SLAC}.
 If the twist-4
contribution to $F_L$ were not proportional to $F_2$,
the twist-4 contribution to $R$ would
depend on the targets.
Using the above fit, we obtain
(at $Q^2=5\ {\rm GeV}^2$)\footnote{Here we have again used the
 LO structure function of ref.\,\cite{GRV}.}
\beq
\int_0^1
F_L^{\tau=4}  dx & = & \frac{1}{2} ( \frac{1}{4}
A^2-\frac{3}{8} A^g )
 \nonumber  \\
 & = &  \int_0^1 8 \kappa^2 F_2^{LT}(x) dx =
 \left\{ \begin{array}{ll}
             0.035  \pm  0.012   &  {\rm GeV}^2 \ \ \mbox{(proton)} \\
             0.023  \pm  0.008  &  {\rm GeV}^2 \ \ \mbox{(neutron)}
                                \end{array} \right.  .
\label{eq12}
\eeq
 As is clear from this expression, the difference between the proton and
the neutron comes only from the difference in  $\int F_2^{LT}dx$.

\section{Constraints on $A^{1,2,g}$}

\vspace{0.2cm}

The experimental data for
 $F_L^{\tau=4}$  (eq.\,(\ref{eq12})) is 2-7 times larger than
those for $F_2^{\tau=4}$ (eq.\,(\ref{eq9})).
Since both $A^1$ and $A^2$ are the matrix elements of the four-quark operators,
their absolute values
are expected to be similar in magnitude.  This together with
 eqs.\,(\ref{eq12}) and (\ref{eq9}) suggests that $A^g$
 at $Q^2=5\ {\rm GeV}^2$
  takes large and negative value to reproduce $F_2$ and $F_L$
 simultaneously. We will come back to this point in section 5.

   From  eqs.\,(\ref{eq12}) and (\ref{eq9}),
 we can  derive two constraints among
$A^1, A^2 $ and $A^g$ :
\beq
A^1 & = & - A^g +
 \left\{ \begin{array}{ll}
             -0.165  \pm  0.061  &  {\rm GeV}^2 \ \ \mbox{(proton)} \\
             -0.093 \pm  0.041  &  {\rm GeV}^2 \ \ \mbox{(neutron)}
                                \end{array} \right.
\nonumber \\
A^2 & = & {3 \over 2}A^g +
 \left\{ \begin{array}{ll}
             0.280  \pm  0.096   &  {\rm GeV}^2 \ \ \mbox{(proton)} \\
             0.184  \pm  0.064  &  {\rm GeV}^2 \ \ \mbox{(neutron)}
                                \end{array} \right.  .
\label{par}
\eeq
The $A^1-A^g$ and $A^2-A^g$ relations  with error bars are
  given as the bands in Fig.\,2.
 The figure shows that it is hard to find a solution where
 $A^{1,2,g}$ are all consistent with zero,
 which clearly indicates
 sizable values of the
 twist-4 matrix elements.  We note that
as long as $A^1$, $A^2$ and $A^g$ do not take too different values among
one another, the typical magnitude of them
 reads 0.1 GeV$^2$ $\sim$ (300 MeV$)^2$
 and a negative value for $A^g$ is favored. (We will
discuss this in detail in section 5.)

 Fig.\,2 gives an useful test of the various nucleon models:
  Any reliable
 models of the nucleon should be able to predict
 the matrix elements within the bands in Fig.\,2.
  One should also note that twist-4 data
 of   $F_3(x)$, although it is not available now,
 will be particularly useful to obtain
 further constraints on $A^{1,2,g}$.

\section{Parameterization of the matrix elements}

\vspace{0.3cm}

Although Fig.\,2 provides us with a model-independent constraint
 among the twist-4 matrix elements, it does not give any definite
 numbers for the matrix elements.
 In this section, we will further introduce a {\rm theoretical}
 assumption to estimate the  magnitude of $A^{1,2,g}$.

\vspace{0.5cm}

\noindent{\bf The Bag Model}

\vspace{0.2cm}

The MIT bag model provides us with the simplest estimate of
 the twist-4 matrix elements. Jaffe and Soldate
calculated $A^1$ and $A^2$ and found that $F_2^{\tau=4}$ in
 the model has an opposite sign
  from  the data (see the footnote 15 of
the  latter reference in \cite{JS}).
Shuryak and Veinstein \cite{SV82} also discussed that
 models without correlation between quarks inside the nucleon
 cannot reproduce the data.
 Let's first generalize the MIT bag model parameterization
to see whether one can remedy the problem
 encountered in \cite{JS}.

The nucleon expectation values of any
  operators in eq.\,(\ref{eq4}) can
be obtained from the bag wave function as follows:
\beq
A^k=\frac{2}{M_N} \int d^3 x \la \hat{p}
 | \op^k_{00}+\frac{1}{3}\op^k_{ii}
 | \hat{p}\ra ,  \,\,\,\,\,k=1,2,g
\label{eq13}
\eeq
where $| \hat{p} \ra $ is the bag state made of three confined quarks.
By using the explicit form of $| \hat{p} \ra$, one obtains \cite{JS},
 $A^1  =  (2/3)f_1 a-(16/9)f_2 a$ and  $A^2 = 2f_1 b
+(16/9)f_2 c$.  Here the factors related to the color-spin-charge read
$a=-16/9 (-4/3)$, $b=-4/3(-8/9)$ and $c=8/9(4/3)$ for the proton (neutron).
 $f_{1,2}$ is related to the spacial wave function of quarks:
 A simple estimate with the bag radius 1 fm
   gives
$f_1=0.0266 \times \alpha_s$ and $f_2=0.0042 \times
\alpha_s$, which leads to
 $A^1 = - 0.018 (-0.014 ) \times \alpha_s  \ \ {\rm GeV}^2$ and
 $A^2 = -0.064 (-0.037) \times \alpha_s  \ \ {\rm GeV}^2$
 for the proton (neutron).
  $\alpha_s$ is the
strong coupling constant and we adopt
 $\alpha_s \sim 0.5$.\footnote{Here it is not
 obvious  whether one should
 use $\alpha_s$ at $Q^2=5\ {\rm GeV}^2$ or something else.
  In this paper, we
 follow the argument in \cite{JS} to estimate an
  ``effective'' value
 $\alpha_s \sim 0.5$.}

The mixed condensate can also be obtained from eq.\,(\ref{eq13})
by assuming abelian electric and magnetic fields.
 The
 electric field vanishes locally within the bag, while
 the magnetic field
 together with the quark wave function in the
 bag gives
\beq
  A^g=  \left\{ \begin{array}{ll}
 0.075  \times \alpha_s   & \ {\rm GeV}^2 \ \ \mbox{(proton})  \\
 0.113  \times \alpha_s  & \ {\rm GeV}^2  \ \ \mbox{(neutron)}.
            \end{array} \right.
\label{16}
\eeq
 Here ${\cal O}^g_{00}$ has a dominant and positive contribution
 to $A^g$.

Adding all the contributions we finally obtain
\beq
 \int_0^1 F_2^{\tau=4} dx =
   \left\{ \begin{array}{ll}
 -0.027 \times \alpha_s   & \ {\rm GeV}^2 \ \  \mbox{(proton)}  \\
 -0.015 \times \alpha_s  & \ {\rm GeV}^2\  \ \  \mbox{(neutron)},
            \end{array} \right.
\label{eq17}
\eeq
and
\beq
\int_0^1 F_L^{\tau=4} dx =
 \left\{ \begin{array}{ll}
             -0.022 \times \alpha_s  & {\rm GeV}^2 \ \ \mbox{(proton)} \\
             -0.026 \times \alpha_s  & {\rm GeV}^2 \ \ \mbox{(neutron)}.
                                \end{array} \right.
\label{eq18}
\eeq
Comparing these with eqs.\,(\ref{eq9}) and (\ref{eq12}),
one finds that the bag model gives incorrect signs
 although the absolute values are the
right order of magnitude.
 The circle (proton) and the cross (neutron) in
Fig.\,2 denote  the prediction of the bag model, which shows that
 the model is inconsistent with the current data.

One may get opposite signs for $A^1$ and $A^2$
 by making $f_2$ comparable to $f_1$.
  However, for any reasonable form
of the wave function, $f_2$ is much smaller
than $f_1$ and in fact the bag model
gives the most generous estimate.   Diquark models give positive
signs for the moments
\,\cite{EF}, but they do not fit the $x$ dependence of the
structure function \cite{SLAC}.

\vspace{1cm}

\noindent{\bf A parameterization based on  flavor structure}

\vspace{0.3cm}

Instead of introducing  more sophisticated
 models of the nucleon, we now
 discuss a different  kind of parameterization motivated by
eq.\,(\ref{eq10}).
Let us first rewrite the matrix elements of
 the operators in eq.\,(\ref{eq4})
 by using the charge operator ${\cal Q}={\rm diag.}(Q_u,Q_d)$,
 \beq
A^1_{p(n)} & = & Q_u^2 K^1_{u(d)}
 +Q_d^2 K^1_{d(u)}-(Q_u-Q_d)^2 K_{ud}^1/2 \ \  ,
 \nonumber \\
A^2_{p(n)} & = & Q_u^2K^2_{u(d)}+ Q_d^2 K^2_{d(u)}\ \  ,
 \nonumber \\
A^g_{p(n)} & = &  Q_u^2 K^g_{u(d)}+Q_d^2 K^g_{d(u)}\ \  ,
\label{eq19}
\eeq
where $K$'s are the matrix elements defined by
\beq
K^i_u & = & {2 \over M^2} \la \bar{u} \Gamma^i_+ \Delta^i_+ u \ra_p,
\,\,\, i=1,2  \nonumber \\
K^g_u & = & {2ig \over M^2} \la \bar{u}    \{ D_+ { },^*F_{+ \mu} \}
 \gamma^\mu \gamma_5  u \ra_p ,\nonumber \\
K_{ud}^1 & = & {2 \over M^2} \la 2
(\bar{u} \Gamma^1_+ u)(\bar{d} \Gamma^1_+ d) \ra_p.
 \nonumber \\
\label{eq20}
\eeq
 Here,
$\Gamma^1_\alpha= \gamma_\alpha \gamma_5 t^a$,  $\Gamma^2_\alpha=
\gamma_\alpha  t^a$ , $\Gamma_+=\frac{1}{\sqrt{2}}(\Gamma_0+\Gamma_3)$
 and
 $\Delta^i_\alpha=\bar{u} \Gamma^i_\alpha u +\bar{d}
 \Gamma^i_\alpha d$ is
 a flavor-singlet operator.
  The neutron matrix elements are obtained from those of the proton by the
isospin symmetry and we have neglected the
strangeness contribution to simplify the analysis.

Noting that the flavor structure of $K_{d}^{1,2,g}$ and
   that of $K_{u}^{1,2,g}$
 are governed by the $d$-quark and the $u$-quark respectively,
 we will introduce an ansatz
 in which the ratio
 $K_d^{1,2,g}/K_u^{1,2,g}$ is equal to the momentum fraction
 of the $d$ and $u$ quarks in the nucleon:
\beq
K^{1,2,g}_d/K^{1,2,g}_u \simeq
 \int x(d(x)+\bar{d}(x))dx/ \int x(u(x)+\bar{u}(x))dx \equiv \beta.
\label{eq21}
\eeq
Here  $u(x), d(x), \cdot \cdot \cdot $ are the usual twist-2 parton
distribution functions. $\beta$ takes a value 0.476 at $Q^2=5$\ GeV$^2$.
  The analogous condition  for $K_d^{2,g}/K_u^{2,g}$ in eq.\,(\ref{eq21})
is a sufficient condition to satisfy
  eq.\,(\ref{eq10}), which can be checked by substituting eq.\,(\ref{eq19})
into eq.\,(\ref{eq12}) and equating the
charge
operators in both sides. Thus essentially it does not bring any new
 constraints.  On the other hand, the condition for $K_d^1/K_u^1$ is
  purely an ansatz:  Although it is plausible from the point of view of the
flavor-structure of the operator, it needs to be checked
 by a non-perturbative method in QCD.

 With eq.\,(\ref{eq21}), we can  reduce the number of matrix elements
  from 6 ($A^{1,2,g}$ for the proton and the neutron)
 to 4 ($K_u^{1,2,g}, K^1_{ud}$).
  Although we have 4 experimental inputs, we cannot determine
 all of them
 uniquely since the ratio of the proton and neutron
 data for $F_L^{\tau=4}$ is
 automatically satisfied in our parameterization.
 Therefore, we will vary
 $K^1_{ud}$  and solve others as  functions
 of $K^1_{ud}$.  We will also limit the
 variation of  $|K^1_{ud}|$ in the range  between
 $|K^1_d|$ and $|K^1_u|$. (In fact, the difference
between ($K^{1}_u$, $K^1_d$) and $K^1_{ud}$ is only the
flavor structure and QCD is flavor-blind, therefore
 these matrix elements should take the
similar values in  magnitude.)
 The resulting values in ${\rm GeV}^2$ unit are given in Table 1.

\vspace{0.5cm}
\cl{
\begin{tabular}{|c|c|c|c|c|c|c|c|}   \hline
  $K_{ud}^1$  &  $K^1_u$  &   $K^2_u$  &  $K^g_u$ & $K_{ud}^1$  &
 $K^1_u$  &   $K^2_u$  &  $K^g_u$
\\ \hline \hline
$K^1_d$ &  -0.173    & 0.203 & -0.238  &
$-K^1_u$ &  0.083    & -0.181 & -0.494  \\ \hline
$(K^1_d+K^1_u)/2$ & -0.112  & 0.110 & -0.300  &
$-(K^1_d+K^1_u)/2$ & 0.112  & -0.225 & -0.523 \\  \hline
$K^1_u$ &  -0.083   & 0.066 & -0.329 &
$-K^1_d$ &  0.173   & -0.318 & -0.585
\\ \hline
\end{tabular} }

\vspace{0.3cm}

\cl{Table 1}

\vspace{0.5cm}

Table 1 gives the following constraints
on the possible range of $A^g$ at  $5\ {\rm GeV}^2$ scale:
\beq
-(540\ {\rm MeV})^2 & < & A^g < -( 340\ {\rm MeV})^2
  \ \ \ \ \ \ ({\rm proton}) \nonumber \\
-(440\ {\rm MeV})^2 & < & A^g < -(280\ {\rm MeV})^2
  \ \ \ \ \ \ ({\rm neutron}) ,
\eeq
which favor the region inside the parallelograms in Fig.\,2.
(Note that  the results in the present
parameterization  are always confined
 inside the bands in Fig.\,2 contrary to those
  of the bag model.)

 Our analysis here suggests that:

\begin{enumerate}

\item As we have discussed in section 4, the matrix element
of the quark-gluon mixed operator $A^g$ is relatively large compared to
the four quark operators at $5\ {\rm GeV}^2$ scale.
 The  magnitude of the former is about  $-(300-500\ {\rm MeV})^2$
 which is consistent with a  typical hadronic scale.
 The sign and the magnitude
 of the matrix elements should be understood
 in a microscopic manner (either
 by lattice QCD or by non-perturbative nucleon models).
   To compare  model calculations with our
  results in a quantitative manner, one needs to
 evolve $A^{1,2,g}$ from 5 GeV$^2$ scale to the typical
 hadronic scale. This requires further knowledge
 of the anomalous dimensions of the operators
 in eq.\,(\ref{eq4}).
   Our result here is also
  relevant to the analysis of
 the QCD sum rules in the nuclear medium\,\cite{HKL}.

\item One can show that $A^1$ and $A^2$ have opposite signs
  from Table 1.
 This causes a relatively strong
cancellation in $F_2^{\tau=4}(x)$ providing with a reason
for the large difference between the data on $F_2$ (eq.\,(\ref{eq9})) and
on $F_L$ (eq.\,(\ref{eq12})).
\end{enumerate}

\vspace{3cm}
S.H.L and T.H. were supported by U.S. Department
 of Energy under grant DE-FG06-88ER40427.
   C. S. and S.H.L  were partly
 supported by Yonsei University Faculty Research
 grant.  Y. K. was supported by the US National
Science Foundation  under grant PHY-9017077.
  One of the authors (Y. K.) thanks W.K. Tung for useful discussions.

\newpage
\cl{{\bf Figure Captions}}

\vspace{1.5cm}

\noindent
{\bf Fig.\,1}: Typical diagrams for the twist-4 contribution to the
forward Compton amplitude: (a) the four quark contribution and (b) the
quark-gluon mixed contribution.

\vspace{1cm}

\noindent
{\bf Fig.\,2}: The twist-4 matrix elements $A_1$ and $A_2$ as a function
 of $A_g$ in the unit of GeV$^2$. They are evaluated at the renormalization
scale $\mu^2 = 5\ {\rm GeV}^2$.
  The band  in solid line (dashed line) is a region allowed
 by the experimental data for the proton (neutron).  The circle (cross)
 is a prediction for the proton (neutron) in the MIT bag model.
 The region inside the parallelogram is allowed in the parameterization
 based on the flavor structure of the twist-4 operators.

\end{document}